# Dissemination stability and phase noise specification of fiber-cascaded RF frequency dissemination

C. Gao[1,2], B. Wang[1,2,*], Xi. Zhu[1,3], Y.B. Yuan[1,3] and L.J. Wang[1,2,3,4,*]

[1] Joint Institute for Measurement Science, Tsinghua University, Beijing 100084, China
[2] Department of Precision Instruments, Tsinghua University, Beijing 100084, China
[3] Department of Physics, Tsinghua University, Beijing 100084, China
[4] National Institute of Metrology, Beijing 100013, China

*Corresponding author: bo.wang@tsinghua.edu.cn

Over the past decade, fiber based frequency dissemination has achieved significant progresses in following aspects. Ultra-stable frequency dissemination over urban fiber link, which can satisfy optical clocks' frequency comparison requirements, has been demonstrated by several groups [1-5]. Optical frequency transfer over 1840 km fiber link [6] makes continental frequency comparison possible. Aiming to construct the time and frequency synchronization network, several frequency dissemination schemes designed for different topological fiber networks have been proposed and demonstrated [7-12]. Based on the disseminated frequency, all of these schemes can be briefly divided into optical frequency dissemination [1-9] and radio frequency (RF) modulated frequency dissemination [10-18]. From the previous works, we can note that almost all of the ultra-long distance (longer than 100 km) frequency dissemination under single-span stabilization mode are realized via optical frequency dissemination. It is because that the linewidth of the disseminated optical frequency is around Hz level. Due to its long coherent length and low chromatic dispersion effect, ultra-long distance dissemination can be realized via simply adding several fiber amplifiers along the fiber route. While for RF modulated frequency dissemination, the laser carrier always has large spectral width. The accompanying chromatic dispersion effect will limit the dissemination distance-long distance transmission will lead to large bit error ratio. So, in fiber-optic communication area, the transfer distance at a certain data rate is often given by the laser's dispersion tolerance. For example, at the 10 GHz modulation rate, the electro-absorption modulated laser's typical dispersion tolerance is 1600 ps. For single mode G652 fiber its dispersion parameter is $D(\lambda)=17\, ps/nm \cdot km$. Consequently, for a 10 GHz RF modulated frequency dissemination using single-span stabilization, the transfer distance is around 80 km. Lower modulation frequency can lead to longer transfer distance, while it cannot achieve high dissemination stability due to its lower phase resolution [13, 18]. Consequently, for longer distance, cascaded frequency dissemination is the best choice, which means the disseminated frequency has to be demodulated and remodulated again using the cascaded frequency stabilization system. Even so, the RF modulated frequency dissemination has its obvious advantages – simpleness, reliability and practicability. It does not require optical frequency comb to connect the transmitting and transmitted signals. The long term continuously frequency dissemination, which gives the dissemination stability at the integration time of $10^6$ s, are only reported using RF modulated frequency dissemination method [19, 20]. The simultaneously time and frequency synchronization is also realized via RF modulated method [15]. These advantages make it more suitable to construct the time and frequency synchronization network which is required in many practical applications, such as deep space navigation (DSN) and radio telescope array.

To study the dissemination stability and phase noise specifications of the fiber-cascaded RF frequency dissemination system, we perform a lab-top experiment using three sets of RF modulated frequency dissemination systems. They are linked by 50 km + 50 km +45 km fiber spools. The dissemination stabilities of each segments and whole system are measured simultaneously. We demonstrate the total transfer stability of the cascaded system can be predict by: $\sigma_T^2 = \sum_{i=1}^{N} \sigma_i^2$ ($\sigma_i$ is the frequency stability of the ith segment, N is the segment number). After that, the phase noise of each segment is also measured and the result shows that the phase noise spectrums of recovered frequency signal can be optimized via narrow band phase lock loop.

i. **Experiments**

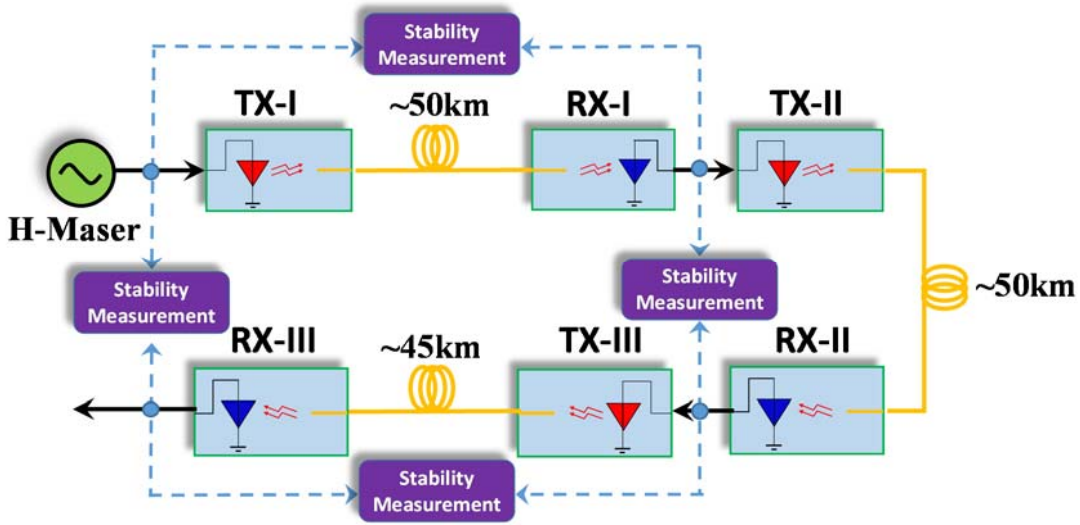

**Figure 1. Schematic diagram of the 3 segments cascaded system.** TX is the stable frequency transmitting module. RX is the receiving module.

The setup and procedure of the cascaded RF modulated frequency transfer experiment is shown in Fig. 1. A 100MHz frequency signal from a synthesizer referenced by the hydrogen maser is used as the reference source of the cascaded system. It is transferred and relayed by 3 segments via cascaded compensation mode. In the transmitting module (TX), the input 100 MHz reference signal is boosted to 9.1 GHz (which leads to a higher signal-to-noise ratio for compensation) using the phase-locking method. After 50 km of transmission in the fiber spool, a stable radio frequency signal was recovered at the receiving module (RX). The module TX-I, RX-I and the 50 km fiber link constitute the first segment of the cascaded system. The recovered signal at RX-I is used as the reference of the next segment. The same structures apply to the other two segments. The frequency dissemination stability of each segment and the whole system are measured simultaneously by four phase comparators. These comparators measure the phase fluctuations between the 100MHz reference signal and recovered 100MHz signal, and the effective sampling bandwidth is 0.5Hz.

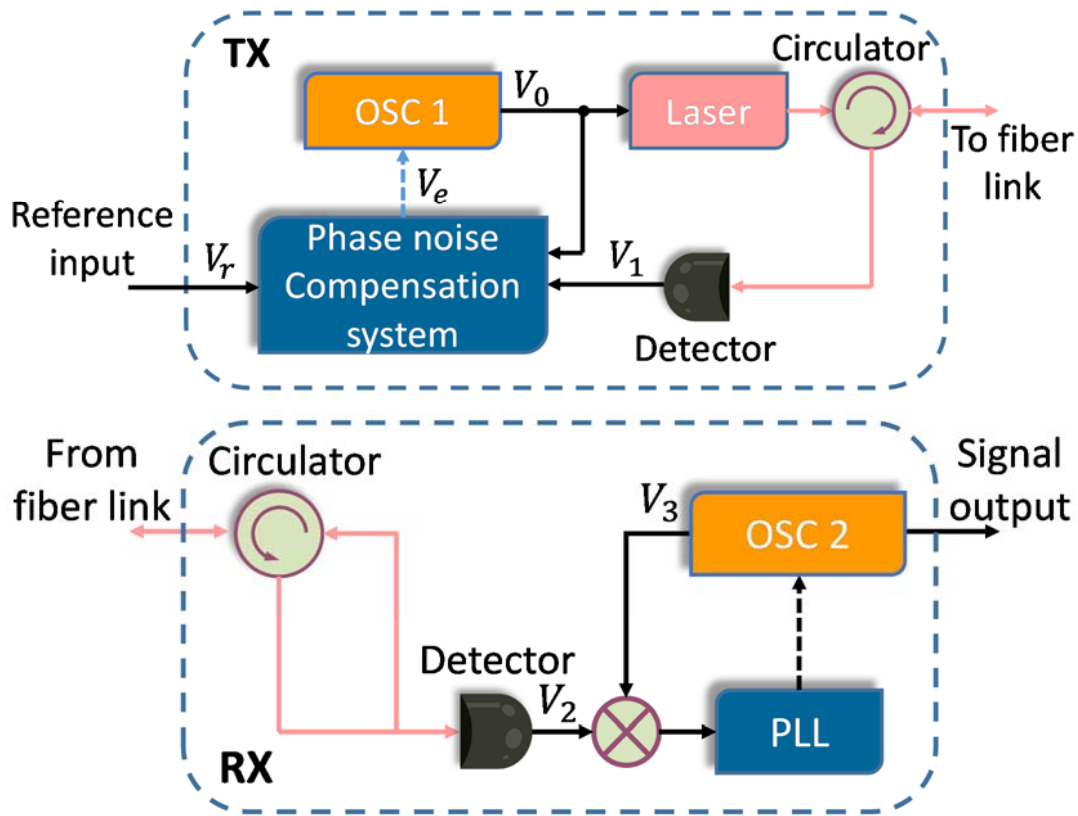

**Figure 2. Simplified schematic of the transmitting and receiving module.** OSC 1 and OSC 2: containing a oven-controlled crystal oscillator (OCXO) and a phase-locked dielectric resonant oscillator (PDRO); PLL: phase locking loop.

The schematic of the TX and RX module (for each segment) is described in Fig 2. In the TX module, the boosted 9.1 GHz reference signal can be expressed as $V_r =$

$\cos(\omega_r t + \varphi_r)$ (without considering its amplitude). A stable oscillator (OSC 1) containing an oven-controlled crystal oscillator (OCXO) and a phase-locked dielectric resonant oscillator (PDRO) generates a signal that can be expressed as $V_0 = \cos(\omega_0 t + \varphi_0)$. Signal $V_0$ is used to modulate the amplitude of the 1550 nm laser carrier and disseminated in the fiber link. After round trip transfer, the 1550 nm laser carrier is demodulated by a detector in the TX. Considering the fiber transmission induced phase fluctuation $\varphi_p$, the rf signal detected at the RX and TX module can be expressed as $V_2 = \cos(\omega_0 t + \varphi_0 + \varphi_p)$, $V_1 = \cos(\omega_0 t + \varphi_0 + 2\varphi_p)$, respectively. According to these three input signals ($V_0, V_1$ and $V_r$), the phase noise compensation system gives the error control signal $V_e$ and feed to OSC 1. When the phase lock loop is closed, its frequency and phase will satisfy the equation:

$$\omega_0 = \omega_r$$
$$\varphi_0 = \varphi_r - \varphi_p. \qquad (1)$$

Consequently, the received frequency signal at the RX site become $V_2 = \cos(\omega_0 t + \varphi_0 + \varphi_p) = \cos(\omega_r t + \varphi_r)$, which is phase locked to that of the reference signal, $V_r$.

Another stable oscillator OSC 2 (also contains an OCXO and a PDRO) at RX module is used as a slave oscillator. One of its output signal is 100MHz, which is used as the reference signal for the next TX module. The basic components of the three TX modules and RX modules are almost the same, except that the three commercial phase locking loop (PLL) used in these RX modules are different.

ii. **Experimental Result**

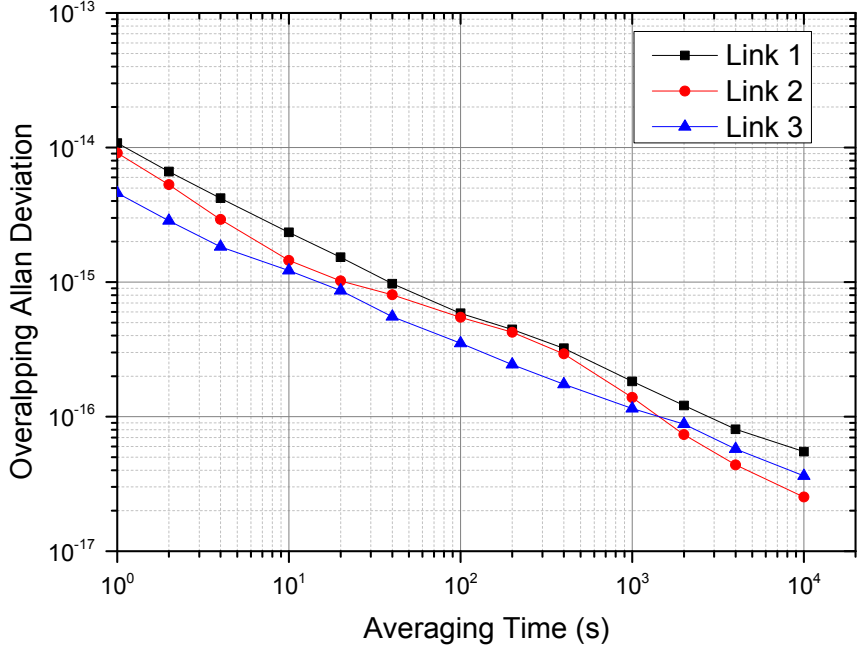

**Figure 3. Frequency dissemination stability of 3 compensated links.** The black line is the frequency dissemination stability of segment 1. The red line is the results of segment 2, and the blue line is the results of segment 3.

By simultaneously comparing the 100MHz reference signal and recovered 100MHz signal's phase, the frequency stability of these three segments is calculated by the computer. The results are shown in Fig. 3.

From the equation of Allan variance, the frequency stability of an oscillator can be calculated as:

$$\sigma_y^2(\tau) = \frac{1}{2M}\sum_{i=1}^{M}(\Delta y_i)^2, \qquad (2)$$

where $\Delta y_i$ is the difference between the (i+1)th and ith of M fractional frequency values averaged over the measurement interval $\tau$: $\Delta y_i = y_{i+1} - y_i$.

For the frequency dissemination stability of the cascaded fiber link, we can considering each segment as an independent oscillator, then the equation (2) is expected to be:

$$\sigma_T^2 = \sigma_I^2 + \sigma_{II}^2 + \sigma_{III}^2 - \frac{1}{M}\sum_{i=1}^{M}(\Delta y_i^I \Delta y_i^{II} + \Delta y_i^I \Delta y_i^{III} + \Delta y_i^{II} \Delta y_i^{III})$$

(3)

The last term, the correlation term, goes to zero as the number of measurements increases; i. e. the compensated phase noise of each segment are independent. Neglecting the correlation term, we can get,

$$\sigma_T = \sqrt{\sigma_I^2 + \sigma_{II}^2 + \sigma_{III}^2}. \qquad (4)$$

Using equation (4) and the measured frequency dissemination stability $\sigma_I$, $\sigma_{II}$ and $\sigma_{III}$, we can get the total frequency dissemination stability $\sigma_T$ of the cascaded system. The calculated $\sigma_T$ and measured $\sigma_{TM}$ are shown in Fig. 4, which is well-matched. In practical system, there are many microwave cables, power splitters, between each TX and RX modules. The phase fluctuations caused by these components are all included in the whole system's stability result $\sigma_{TM}$. While these phase fluctuations are not contained in the frequency stability measurement of each segment, so the measured $\sigma_{TM}$ is a little worse than the calculated $\sigma_T$.

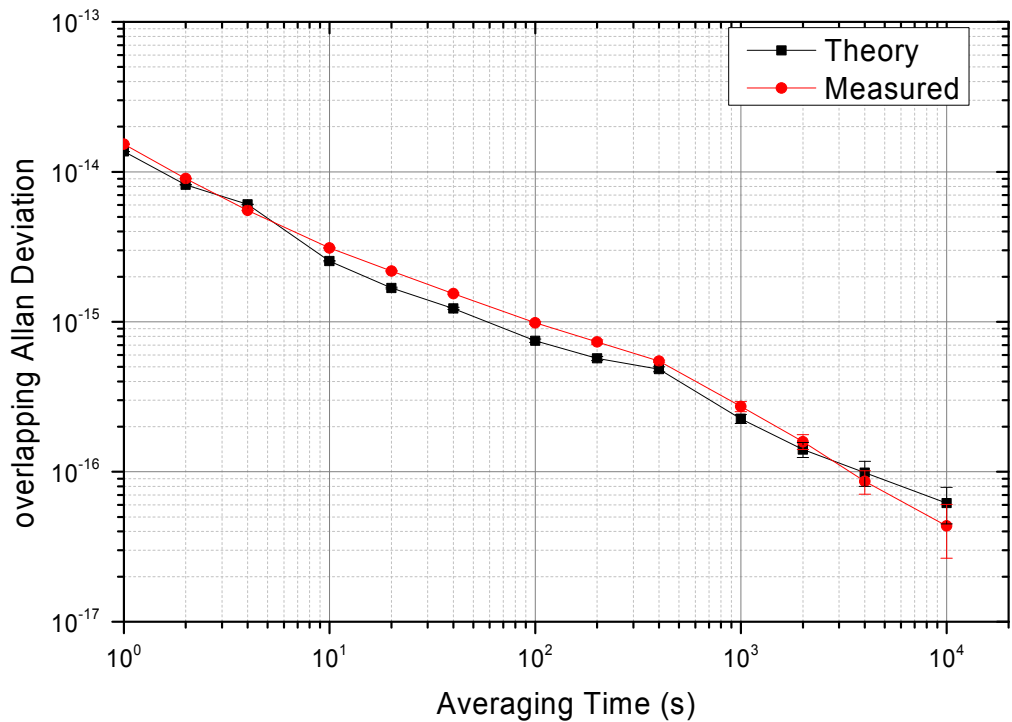

**Figure 4. The measured and calculated frequency stability of the cascaded system.**

The phase noise of the cascaded system is measured by a universal phase noise analyzer (E5052B, Agilent). The results are shown in Fig. 5. The black line is the phase noise of the hydrogen maser referenced synthesizer (100MHz). The red line is the phase noise of the 100MHz frequency signal recovered at RX-I. Here, peak 1 at about 300Hz is caused by the fiber loop delay. Peak 2 at about 1 kHz, is caused by the servo loop's bandwidth of RX-I. The blue line is the phase noise of the 100MHz frequency signal recovered at RX-II. The only peak on the blue line is caused by the servo loop used here. For in RX-II, the servo loop's bandwidth is about 100Hz, which is smaller than the loop delay caused bandwidth. The green line is the phase noise spectrum of the 100MHz frequency signal recovered at RX-III. Peak 4 at 100Hz is the phase noise inheriting from RX-II. Peak 5 is caused by fiber loop delay between TX-III and RX-III. Peak 6 is caused by the servo loop's bandwidth at RX-III. For each segment, the recovered signals' phase noise spectrum at high frequency domain are different, but at low frequency (<10Hz) they are all strictly locked with the reference signal. Which means at low frequency offset the frequency dissemination system introduces negligible noise compared to the source.

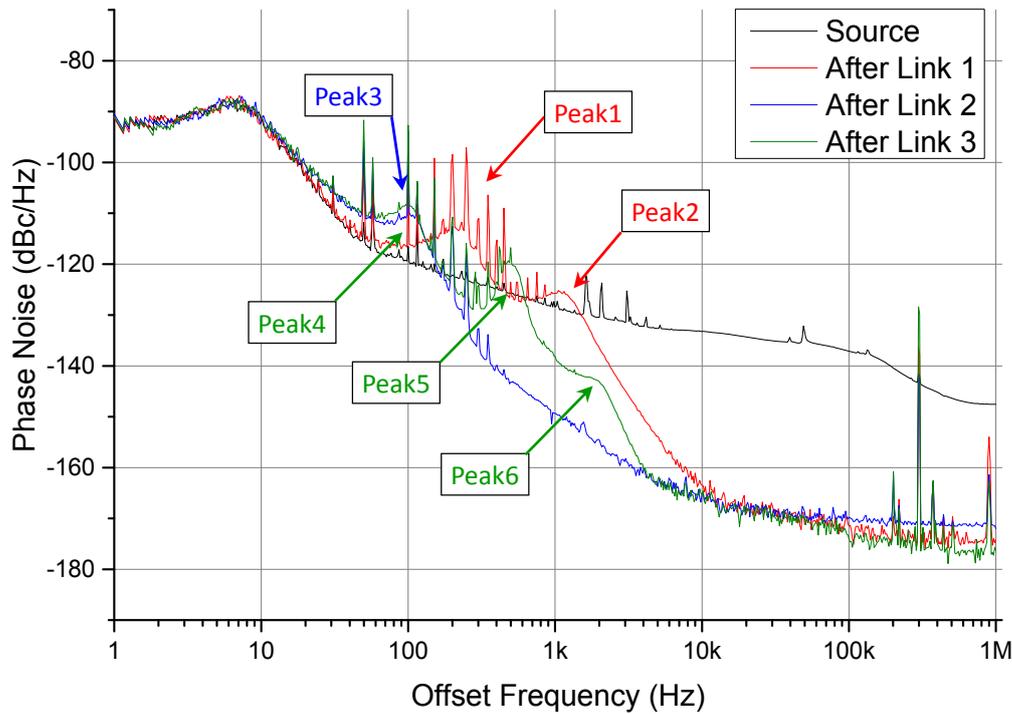

**Figure 5. The phase noise of the cascaded system.** The black line is the phase noise of the reference signal. The red line is the phase noise of the signal recovered by RX-I. The blue line is the phase noise of the signal recovered by RX-II. The green line is the phase noise of the 100MHz signal recovered at RX-III.

For many applications, such as the square kilometer array telescopes and deep space network, they not only need the long term frequency stability, the short term phase noise of the signal is also very important. Consequently, we use a narrow band clean up oscillator at the RX module, which can highly reduce the phase noise caused by the feedback loops. Fig 6 is the phase noise of these RX modules' output after adding a homemade narrow band clean up oscillator. The green line is phase noise spectrum of the clean up oscillator. The homemade clean up oscillator's bandwidth can be tuned and set based on the phase noise specification of free-running OCXO and H-Masr clock. In our case, the PLL's bandwidth is set as 10 Hz. The phase noise after the clean up oscillator is coincide with that of H-maser clock below 10 Hz frequency offset, and is even better after 10Hz frequency offset. The phase noise of the output signal is significantly improved.

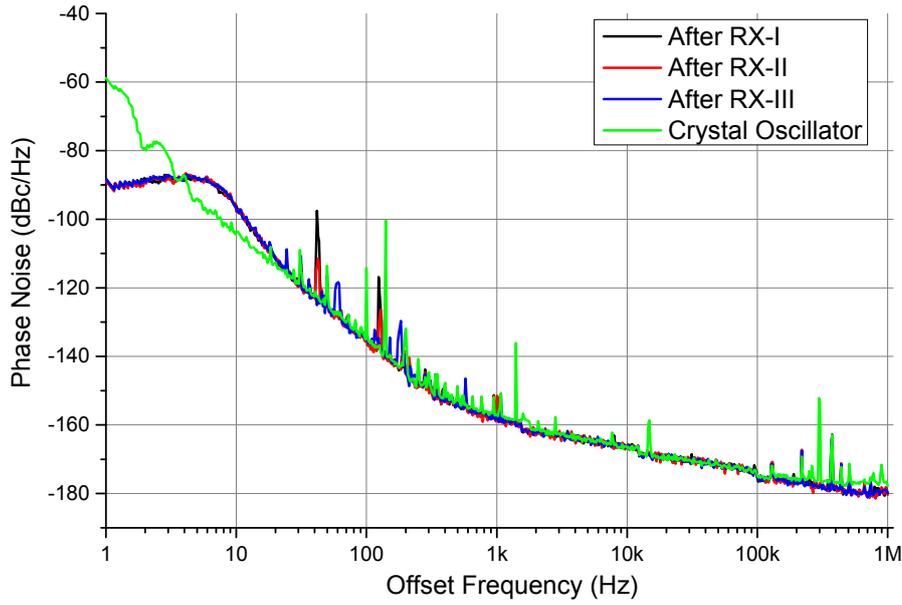

**Figure 6. The phase noise of these RX modules' output signals with narrow band clean up oscillator.** The Green line is the phase noise of the free running crystal oscillator.

### iii.     Conclusion

We demonstrate a RF transfer over a 145 km fiber link by a cascaded system with three segments. The frequency dissemination stability of each segment and the total fiber link are measured simultaneously. The frequency dissemination stability of the total link is $1.3 \times 10^{-14}$/s, $4.05 \times 10^{-17}/10^4$s. From the measured and calculated dissemination stability results, and the transfer stability of the cascaded system can be predict by: $\sigma_T^2 = \sum_{i=1}^{N} \sigma_i^2$ ($\sigma_i$ is the frequency stability of the ith segment, N is the segment number). If the stability of each segment is similar, the total stability (in Allan Deviation) degrades by only a factor of $\sqrt{N}$, which would not degrades dramatically during long distance transition. Furthermore, the phase noise of the recovered 100 MHz signal is improved by a homemade narrow bandwidth clean up oscillator, which is coincide with that of H-maser clock below 10 Hz frequency offset, and is even better after 10 Hz frequency offset.